\documentclass{article}  

\usepackage{latexsym}
\usepackage{graphicx}
\usepackage{psfrag}

\usepackage{pslatex}

\begin{document}

\title{Beyond Quantum Field Theory:\\ 
Chaotic Lattices?}

\author{ 
  Tam\'as S. Bir\'o$^1$, Berndt M\"uller$^2$ and Sergei Matinyan$^3$
}

\maketitle

\newcommand{\address}[1]{
   
   \begin{enumerate}
   \item[]\rm\raggedright #1
   \end{enumerate}
}

\begin{center}
\address{$^1$ KFKI Research Institute for Particle and Nuclear Physics 
H-1525 Budapest Pf.49, Hungary }
\address{$^2$ Physics Department, Duke University 
Durham, NC 27708, USA } 
\address{$^3$ Yerevan Physics Institute, Yerevan, Armenia}
\end{center}



\newcommand{\va}{\vspace{0mm}}
\newcommand{\vb}{\vspace{5mm}}


\newcommand{\be}{\begin{equation}}
\newcommand{\ee}{\end{equation}}
\newcommand{\ba}{\begin{eqnarray}}
\newcommand{\ea}{\end{eqnarray}}
\newcommand{\NL}{\nonumber \\}

\newcommand{\binom}[2]{ \left( \begin{array}{c} #1 \\ #2 \end{array} \right)}
\newcommand{\E}[1]{ \langle #1 \rangle }
\newcommand{\ov}[1]{ \overline{#1} }

\begin{abstract}{
We review the idea of chaotic quantization, based on the chaotic dynamics
of classical lattice gauge systems as well as on non-abelian plasma physics
in the infrared limit. The basic conjecture between Planck constant and
properties of a five-dimensional lattice ($\: \hbar = aT$) is demonstrated
numerically for the U(1) lattice gauge group.
}
\end{abstract}



\section{Chaotic quantization}\label{ChQ}

\va
By promoting the idea of chaotic quantization 
we intend to call the attention to a possibility
which could resolve at least part of the frustration stemming
from the fact that gravity is resisting theoretical intentions
to be quantized. String theory, while it approaches the problem
on the basis of quantum field theory not only requires 10
space-time dimensions to be consistent, but also calls upon
the additional concept of compactification in order
to arrive at standard model particle physics.

\va
Chaotic quantization has been suggested by us\cite{bib1,bib2} 
as an opposite
strategy, starting with classical gauge field theory which
due to its own Hamiltonian dynamics is inherently chaotic \cite{bib3,bib4}.
It evolves ergodically - enough time given - 
in field-configuration phase space
and this process leads in the infrared limit to a stationary distribution
of lower dimensional sub-configurations. These, as result
of the higher dimensional {\em classical} dynamics, are distributed
as {\em quantum} field theory requires in the imaginary time formalism.
The chaotic quantization is
a particular form of the stochastic quantization \cite{bib5}, it works 
self-contained, not requiring any assumption of an external heat bath
or noise.

\va
One should in principle
combine this mechanism with the interesting results of C. Beck 
\cite{bib6,bib7,bib8},
who noted that chaotic maps closest to the most general white noise
assumption of the stochastic quantization, namely Tchebisheff polynoms,
offer a classification of possible  chaotic quantization outcomes.
In his scheme he was able to derive ratios of the most important
Standard Model parameters, like leptoquark masses and relative
coupling strengths, as stable zeros of the effective potential
for a higher dimensional lattice system developing under a chaotic
dynamics. He actually predicted neutrino masses before
any measurement pointing towards a mass square difference
between neutrino generations (which implies that all neutrinos
cannot be massless) \cite{bib9}. 

\va
Set into the above perspective the chaotic quantization mechanism
is worth to study. As a working example we take a 5-dimensional lattice
on which a classical, pure U(1) gauge theory resides. 
This simple group still shows chaotic dynamics on finite lattices
\cite{bib10}.
Although in the continuum limit U(1) is not chaotic, 
at finite lattice spacing and strong coupling classical chaotic dynamics 
was already demonstrated in the 3-dimensional case. This behavior
is probably correlated with the presence of magnetic monopole
anti-monopole pairs \cite{bib11,bib12}.

\va
Our main conjecture between the 5-dimensional classical
theory with chaotic dynamics and the 4-dimensional quantum field
theory can be comprised into the following formula between
the 'normal' (4-dimensional) Planck constant and two physical
characteristics of the higher dimensional theory, its temperature
and lattice spacing:

\be
 \hbar = a T 
\ee

Usually such a formula is read in an opposite direction, automatically
relating a Planck mass ($M_P=T$) with a Planck length ($\ell_P=a$).
Our philosophy here, however, views $\:\hbar$ as a constant of
nature factorized to two other, underlying properties of the 
(in the present theory 5-dimensional) world. An analogy of this
situation is the classical electrodynamics formula factorizing 
another constant of nature, the speed of light:

\be
1/c^2 = \epsilon_0 \mu_0.
\ee

Taking $c=const.$ as a postulate we arrive at the theory of
special relativity, one derives the laws of Lorentz transformation
in the framework of mechanics without making any reference to
electric or magnetic fields (as it has been shown by Albert Einstein).
Maxwell theory on the other hand, as a classical field theory
regards $\epsilon_0$ and $\mu_0$ as independent properties of the
physical vacuum, as dielectric constant and magnetic permeability.
Light waves are solutions of Maxwell theory and the speed of light
is calculable. The relation between nowadays Quantum Field Theory
and an underlying classical field theory is analogous to this.
Furthermore, as ether does not need to exist for Maxwell theory
to work, the five dimensional lattice also may prove to be just
a theoretical construct without measurability even at the Planck scale.


\section{Chaos in gauge theory}\label{ChGT}

\begin{figure}[ht]
\begin{center}
\includegraphics[width=0.25\textwidth,angle=-90]{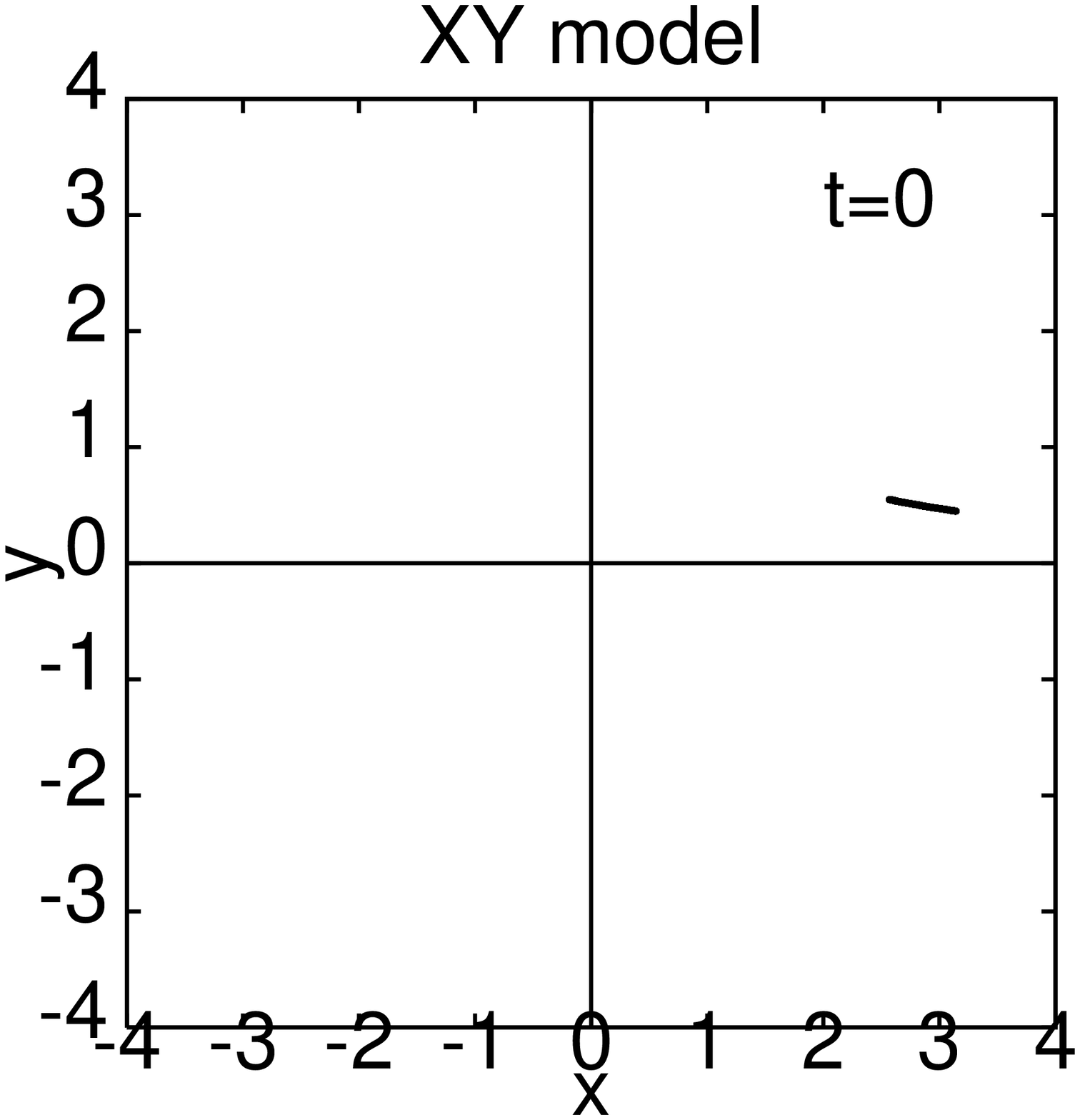}\includegraphics[width=0.25\textwidth,angle=-90]{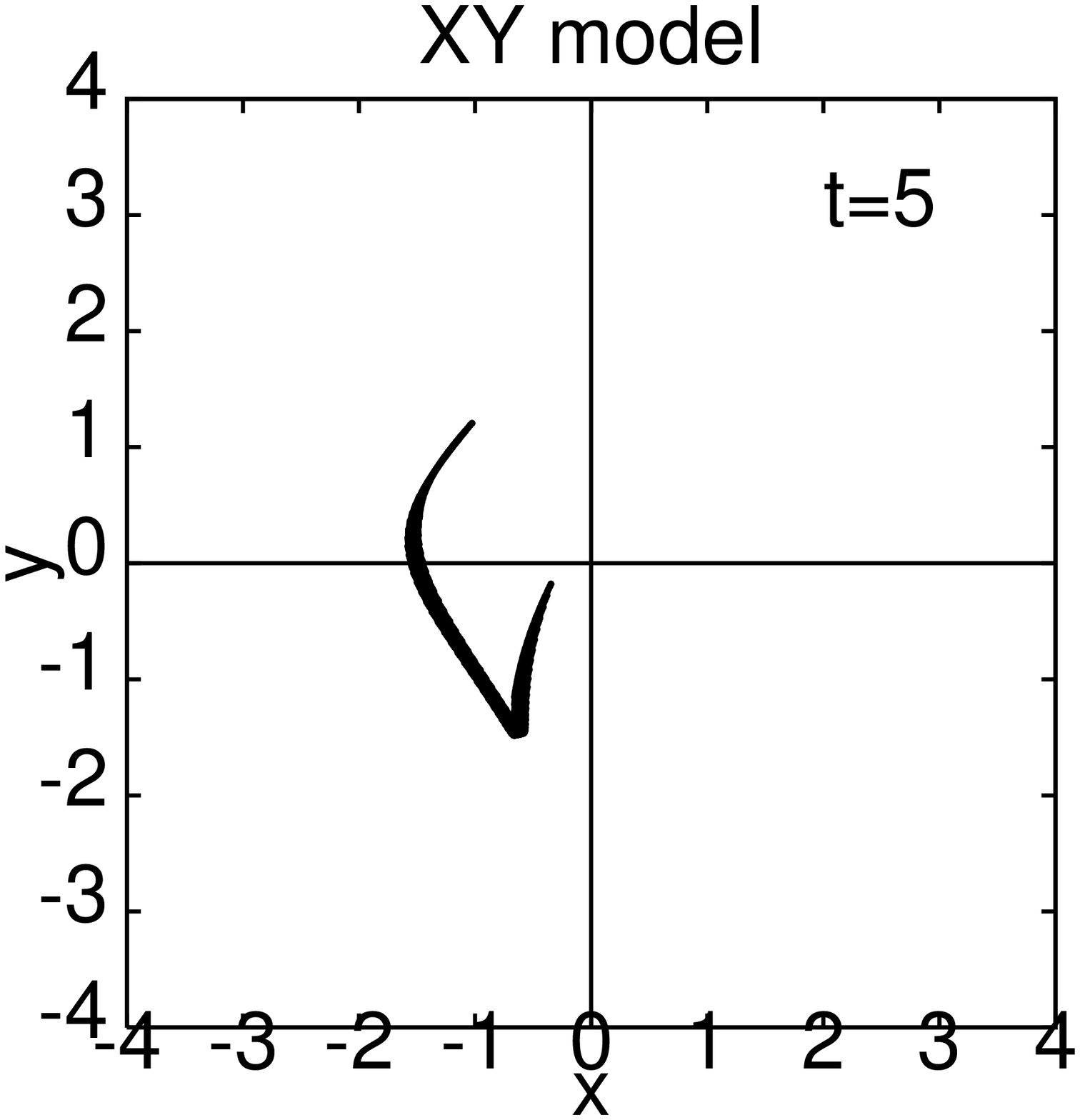}\includegraphics[width=0.25\textwidth,angle=-90]{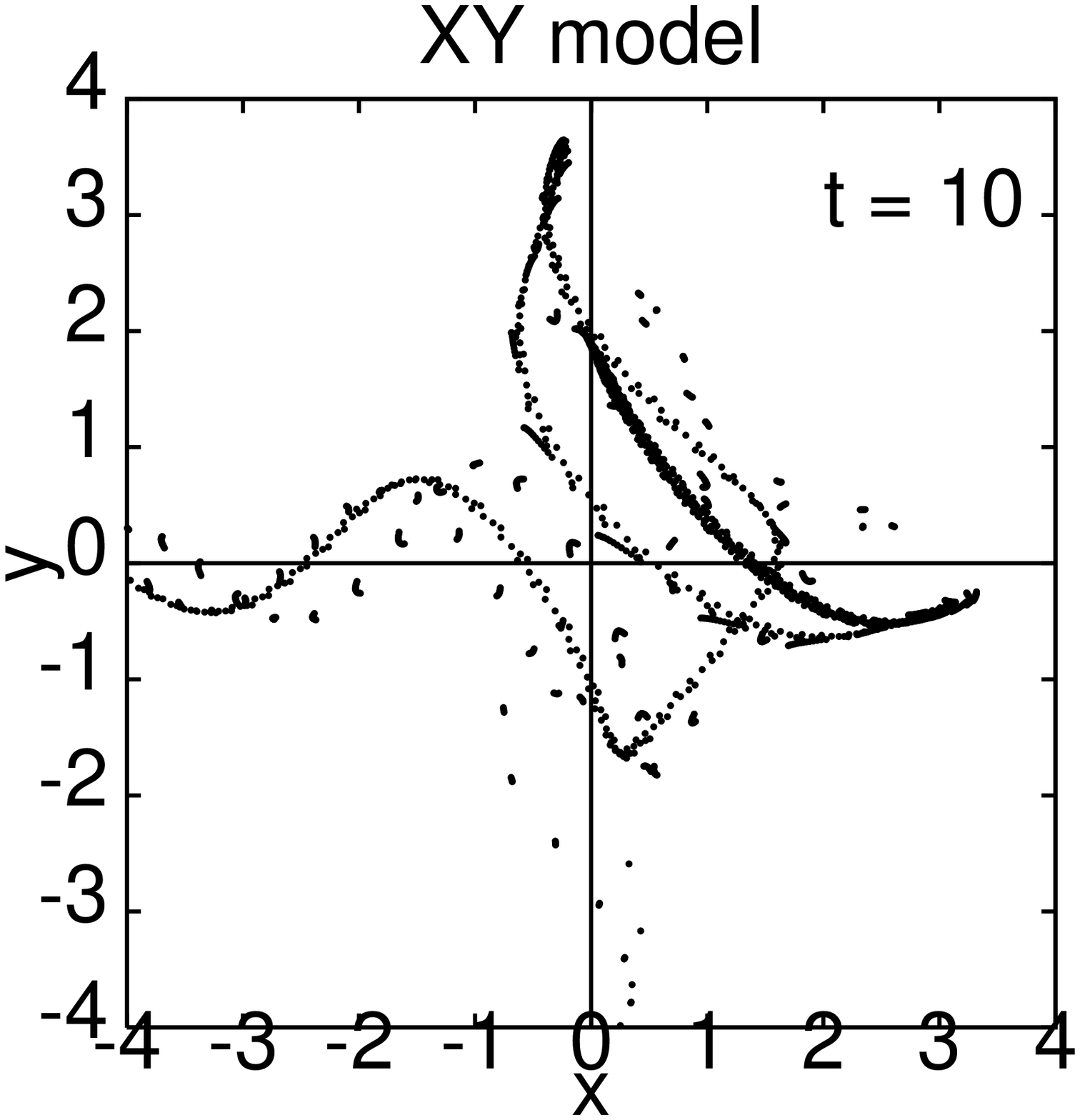}

\vspace{7mm}

\includegraphics[width=0.25\textwidth,angle=-90]{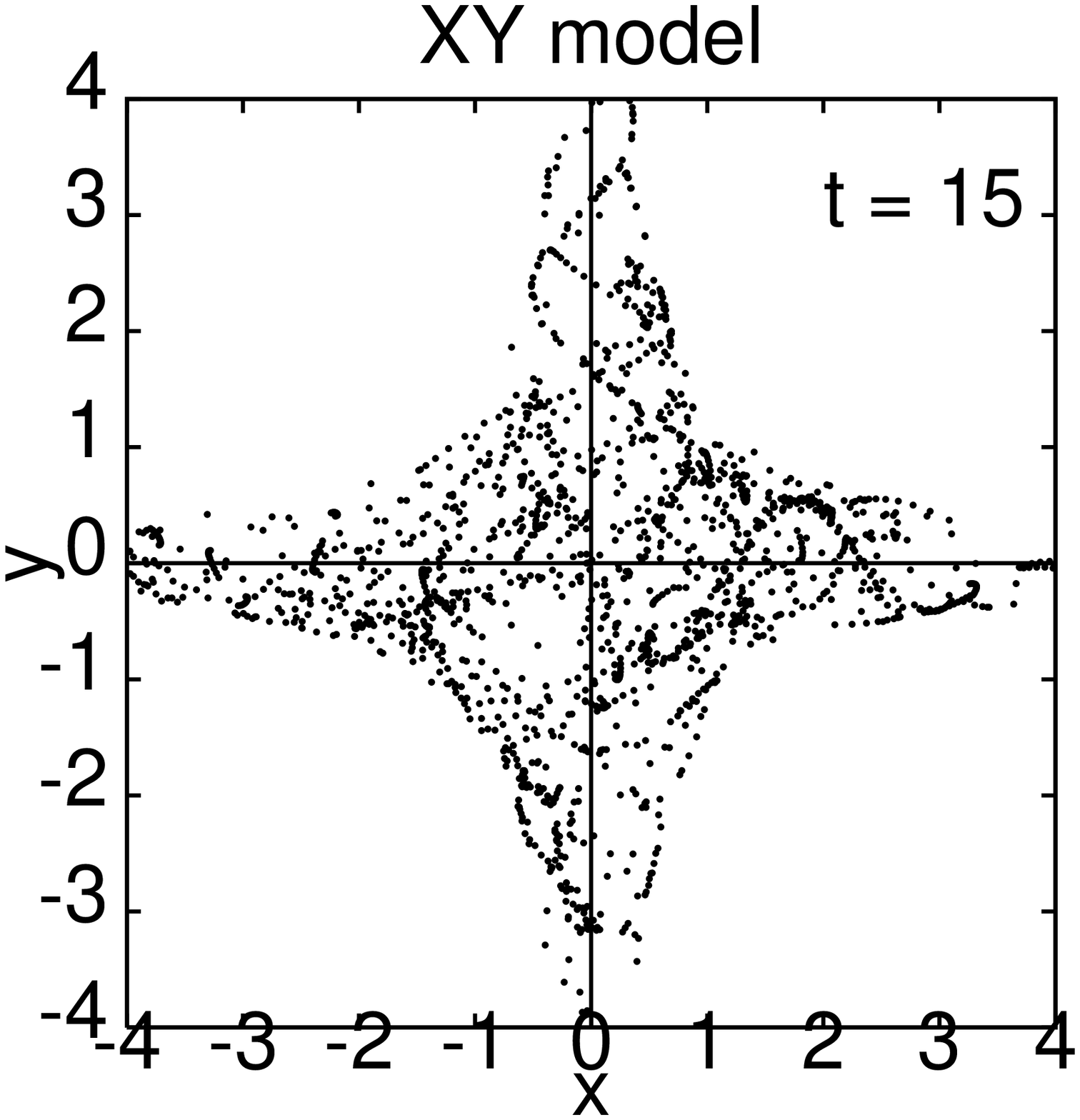}\includegraphics[width=0.25\textwidth,angle=-90]{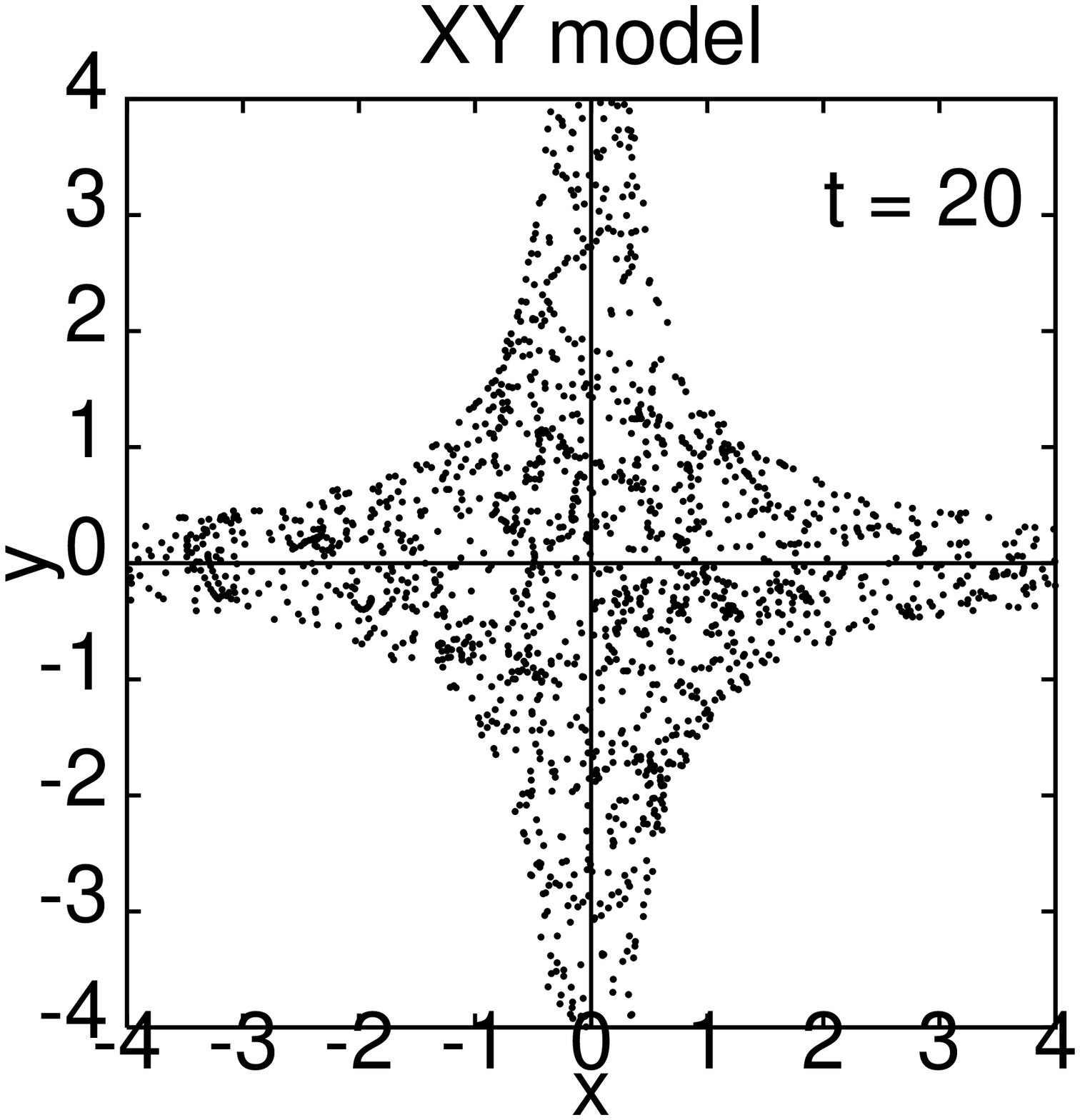}\includegraphics[width=0.25\textwidth,angle=-90]{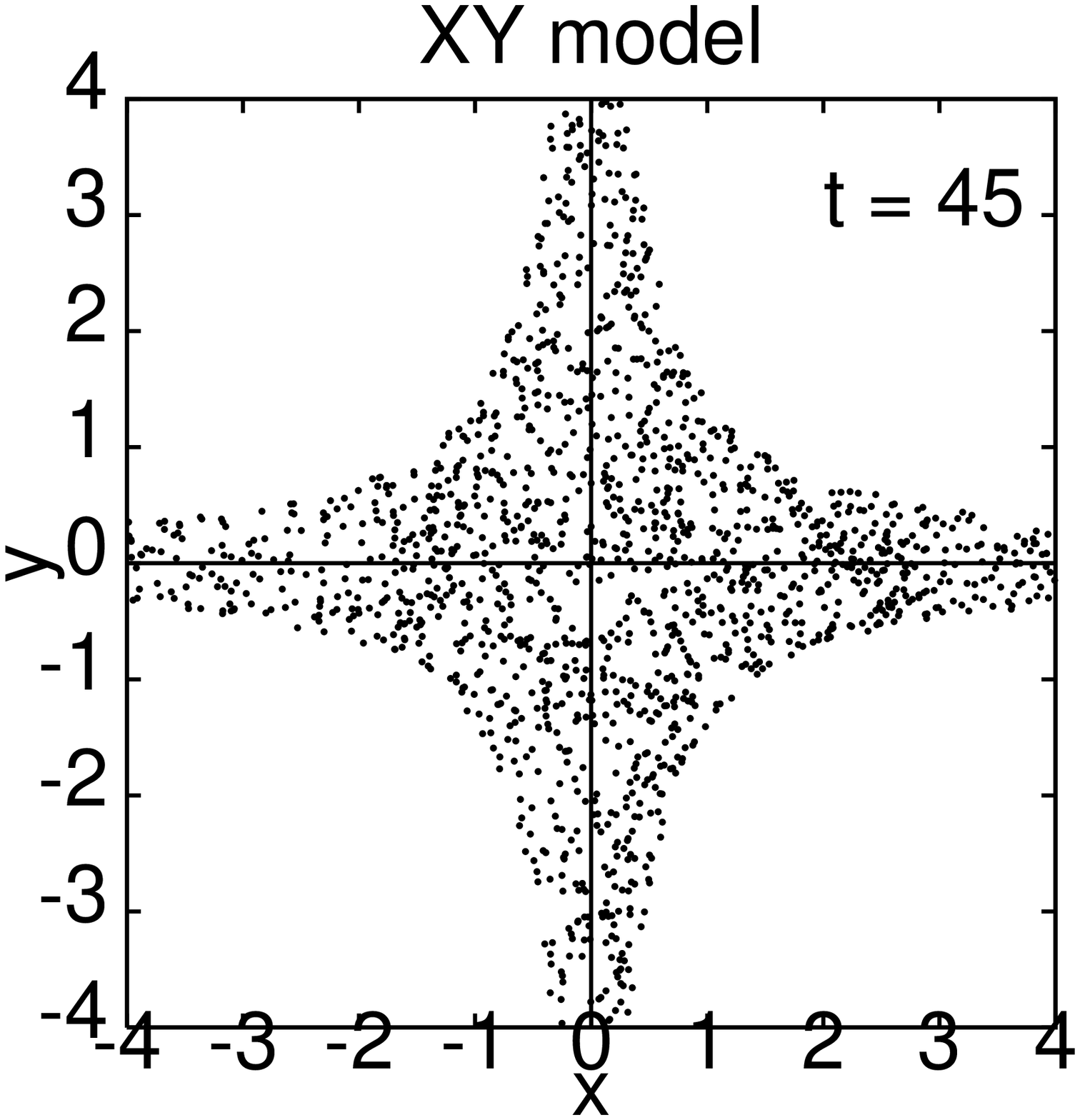}

\end{center}
\caption{
 Time evolution of initially adjacent trajectories in the $xy$-model.
 One hundred initially adjacent points eventually scatter over the
 whole classically allowed phase space.
}
\label{Fig0}
\end{figure}

\va
First non-abelian then also abelian gauge theories has been 
studied with respect to chaotic behavior. In the eighties model
systems, with a few, selected degrees of freedom with long wavelength
has been studied. The most characteristic results stem from SU(2)
Yang-Mills theory, considering two $k=0$ modes of different
polarization and color. Due to the simple Hamiltonian,
\be
H = \frac{1}{2} \left( \dot{x}^2 + \dot{y}^2 \right) +
\frac{1}{2}g^2 \, x^2y^2,
\ee
this $xy-model$ \cite{bib13} 
was studied very often. It is a classically chaotic
system with a null-measure of regular periodic orbits; 
the turning points of the classical motion lie on
hyperboles defocusing so any nearby, parallel trajectories
(cf. Fig.\ref{Fig0}).

\va

\va
Numerical studies on lattices followed in the 1990-s. Mainly the gauge
groups U(1), SU(2) and SU(3) has been studied, all showed chaotic
behavior \cite{bib14,bib15,bib16}. 
The scaling of the leading Lyapunov exponent with the
scaled total energy of the classical lattice, 
$ a \lambda_{{\rm max}} \sim ag^2H $ 
for non-abelian gauge groups, as well as the extensivity of
the Lyapunov spectrum and the Kolmogorov-Sinai ($KS$) entropy -- related
to the ergodization speed -- were studied \cite{bib17}. 
Interpreting the KS-entropy
as the physical entropy of the lattice gauge system even an attempt
can be done to extract a classical equation of state \cite{bib18}.

\va
A basic interest lies in the investigation whether classically
chaotic systems and configurations are also special for the quantum
pendant. Correlation between chaos and confinement has been observed
in SU(2) and U(1) lattices comparing quantum Monte Carlo results
and classical Hamiltonian dynamics in 3+1 dimensions \cite{bib19}. 
For the U(1)
group in the strong coupling phase a strong tendency could have
been observed between the presence of magnetic monopole anti-monopole
pairs and the chaotic behavior \cite{bib11}.

\va


\section{Higher dimensional plasma physics}\label{PLASMA5}

\va
The main point of the stochastic quantization \cite{bib5,bib21}, 
namely generating
field configurations distributed according to Boltzmann weights,
\be
 {\cal P} = \exp \left( -\beta S_E[A'] \right)
\ee
for pure gauge theories, where the original vector potential
variable, $A$ is rescaled to $A'=gA$ and as a consequence
$\beta = 1/g^2\hbar$ due to the scaling of the action $S[A]=S[gA]/g^2$.
These Boltzmann weights occur in Feynman path integrals when
calculating expectation values in quantum field theory. At the same
time they can be regarded as a stationary distribution of a
corresponding Fokker-Planck equation,
\be
\sigma_5 \frac{\partial}{\partial t_5} {\cal P} =
\int \! d^4x \, \frac{\delta}{\delta A}
\left( \frac{1}{\beta} \frac{\delta}{\delta A} {\cal P} + \frac{\delta S}{\delta A}
{\cal P} \right). 
\ee
Formally $1/\beta$ can be interpreted as a temperature of the five dimensional
system, $T_5$. The Fokker-Planck equation is equivalent with solutions of
corresponding Langevin equation,
\be
 \ddot{A} + \frac{\delta S}{\delta A} = J
\ee
where the source current density is split to a dissipative term 
(like ohmic resistance) and to a fluctuative (noise) term:
\be
 J = -\sigma \dot{A} + \zeta.
\ee
This noise is usually treated simplified, as a Gaussian white 
noise with zero
mean and a correlation sharply localized in space and time:
\be
 \langle \zeta(x_1) \zeta(x_2) \rangle = 2\sigma T \delta^{4}(x_1-x_2).
\ee
In the infrared limit the low frequency components dominate the relevant
vector potential configurations, the radiative term $\ddot{A}$ can be neglected
besides the dissipative term $\sigma \dot{A}$. The typical frequency is
small, $\omega \ll \sigma$, the typical time is large, $\tau \gg 1/\sigma$
\cite{bib22}.
The effective model to electrodynamics is a Langevin plasma described by
\be
 \sigma \dot{A} + \frac{\delta S}{\delta A} = \zeta.
\ee
This equation is analogous to the well-known Brown motion, the mean features
of its solution, too: the action $S$ after long enough time is distributed
around its ergodic limit of $T/2$ and initial correlations decay 
exponentially
with the characteristic time of the corresponding damping constant,
$t_{{\rm char}}=1/\gamma=\sigma/|k|^2.$ The long time average of such
correlations with the initial value are interpreted in the lower 
dimensional field theory as propagators.

This analogy with plasma physics in the continuum limit led us to detailed
investigations of dynamical time scales: the thermal ($\hbar/T$), electric
($\sqrt{\hbar}/gT$) and magnetic screening length ($1/g^2T$) in a usual,
three dimensional plasma depend on the temperature, on the Planck constant
and on the coupling constant of the original gauge theory, $g$.
In the long time limit the plasma dynamics leads to distributions
simulating a dimensionally reduced, three dimensional field theory with
the effective coupling $g_3=g^2T$. It can happen only for pure gauge theories,
with mass scale invariance. The magnetic part of the energy is identical with
a lower dimensional Maxwell or Yang-Mills action:
\be
 \frac{1}{2}B_iB_i = - \frac{1}{4} F_{ij}F^{ij}.
\ee
The effective approach with white noise and Langevin equation is valid
for the long-time behavior, $t \gg 1/g^2T$.
Table 1 summarizes the most important features of traditional plasma physics,
in the third column showing the corresponding formulae for classical
lattices. Since most of the time constants in plasma physics are coefficients
in a linear response approach, they can be calculated on a classical lattice
as well as in quantum field theory. Here no Planck constant occurs, but the
lattice spacing plays a basic role. These two approaches coincide if
the universal relation $\: \hbar = a T$ is assumed.


\begin{small}
\begin{table}
\begin{center}
\begin{tabular}{|c|c|c|}
\hline 
 & 3-dim QFT plasma & 3+1 class. lattice \\ 
\hline  
\hline
$d_m$ & $1/g^2T$ & $1/g^2T$ \\ 
\hline
$d_e$ & $\sqrt{\hbar}/gT$ & $\sqrt{a/g^2T}$ \\ 
\hline
$\omega^2$ & $g^2T^2/\hbar$ & $g^2T/a $ \\ 
\hline
$ \gamma$ & $ g^2T $ & $g^2T$ \\ 
\hline
$\sigma $ & $T/\hbar$(log) & $1/a$(log) \\ 
\hline
$d_m \gg d_e$ & $g^2\hbar \ll 1$ & $ a \ll 1/g^2T$ \\ 
\hline
\end{tabular}
\end{center}
\caption{
 Scales in plasma physics and on the lattice
 show a universal scaling: $\hbar = aT$.
}
\end{table}
\end{small}


\begin{small}
\begin{table}[ht]
\begin{center}
\begin{tabular}{|c|c|c|}
\hline 
 & d dim. QFT plasma & d+1 class. lattice \\ \hline 
\hline
$d_m$ & $\frac{\hbar}{G^2} = \frac{\hbar^{d-3}}{g^2T^{d-2}}$ & 
	$\frac{a}{G^2} = \frac{a^{d-3}}{g^2T}$ \\ 
\hline
$d_e$ & $\frac{\hbar}{GT} = \frac{\hbar^{d/2-1}}{gT^{(d-1)/2}}$ & 
	$\frac{a}{G} = \frac{a^{d/2-1}}{g\sqrt{T}}$ \\ 
\hline
$\sigma $ & $\frac{d_e^{d-5}}{g^2T}$ & $\frac{d_e^{d-5}}{g^2T}$\\ 
\hline
$ g_d^2$ & $ \frac{g^2T}{\hbar} $ & $ \frac{g^2}{a}$ \\ 
\hline
$ G^2 \ll 1$ & $g^2T^{d-3}\hbar^{4-d}$ & $ g^2T a^{4-d} $ \\ 
\hline
\end{tabular}
\end{center}
\caption{
 Scales in plasma physics and on the lattice in arbitrary dimensions.
 All formulae coincide if $\hbar = a T $.  
Here $g$ is the original coupling, 
$G$ the weak parameter signaling the infrared limit
and $g_d$ the effective coupling of the dimensionally reduced theory.
}
\end{table}
\end{small}
These calculations can be repeated in arbitrary number of 
Euclidean dimensions,
including the 4+1 - dimensional case of traditional stochastic quantization.
Table 2 shows the characteristic results.

\section{U(1) lattice model}\label{U1}
 
\va
In this section we report about numerical simulations on a U(1)
lattice gauge system both in 4 and 5 dimensions. The former
was simulated by quantum Monte Carlo techniques in order to
reproduce long known standard results \cite{bib23,bib24}, 
the latter independently
by 4+1-dimensional classical Hamiltonian dynamics known to be
chaotic from our former studies. In both cases regular (rather
small, $4^4$) lattices are considered.

\va
In order to appreciate the computational complexity one faces to,
we review briefly basic formulae and techniques of lattice gauge
theory calculations.
In these models of continuum field theory (both in the classical
and quantized version) lattice links 
starting at point $x$ and pointing in the $\mu$ direction are associated
with phases, $A_{\mu}(x)$ of unimodular complex
numbers -- elements of U(1) -- $U=\exp(igaA_{\mu}(x))$, 
while the lattice action
is constructed from phase sums around elementary plaquettes,
upon using lattice forward derivatives ($a\partial_{\mu}f = 
f(x+ae_{\mu}) - f(x)$). The plaquette phase sums satisfy
\be
F_{\mu\nu}(x) = \partial_{\mu}A_{\nu}(x) - \partial_{\nu}A_{\mu}(x),
\ee
and determine the lattice action
\be
S = \frac{1}{g^2} \sum_x \sum_{\mu > \nu} (1 - \cos(ga^2F_{\mu\nu})).
\ee
Here the summation runs over all lattice plaquettes in planes 
each characterized by a pair of two (ordered and different) 
direction indices, $\mu > \nu$, and
attached with its corner to the site $x$. In the continuum limit
$a \rightarrow 0$ the action of the classical electrodynamics is
recovered. Quantum Monte Carlo algorithms produce and sample
$U$ lattice link values (so called configurations) which are weighted
by a Boltzmann type factor
\be
w = e^{-S/\hbar}.
\ee
This corresponds to the evaluation of Feynman path integrals in
quantum field theory in the imaginary time formalism.

\va
The classical Hamiltonian approach on the other hand 
uses the Hamiltonian split to electric and magnetic parts,
\be
H = \sum_{x,\mu} \frac{a}{2g^2} |\dot{U}|^2 + E^{{\rm magn}}[U]. 
\ee
Here the magnetic contribution to the total Hamiltonian,
\be
E^{{\rm magn}}[U] = \frac{1}{g^2a} \sum_x \sum_{i<j}
\left( 1 - \cos(ga^2 F_{ij}) \right),
\ee
is a sum over plaquettes lying in spatio-spatial (hyper)planes.
This sum is formally equivalent to the Euclidean action of the
same lattice gauge theory in one dimension lower. This self-similarity
of pure gauge actions is an essential ingredient for the particular
mechanism of chaotic quantization we are pursuing now.

\va
We present results of numerical simulations of a five
dimensional classical Hamiltonian U(1) lattice system 
and compare its evolution
in the 5-th coordinate with traditional quantum Monte Carlo
generated configurations on a four dimensional lattice, using
the four dimensional U(1) lattice action.
In the classical Hamiltonian approach
the evolution of the $U$ configurations proceeds in a 5-th dimension,
often called 'fictitious' time when it has been used as a method
for stochastic quantization. The important difference is,
that sofar always an external heat bath or a white noise for solving
Langevin type equations has been added to the evolution; we consider
here pure classical Hamiltonian dynamics with no other source of noise
or fluctuations.

\va
As by construction $aE_5^{{\rm magn}} = S_4$ 
(since the dimensionless plaquette sum 
or average over the 4-dimensional (sub)lattice
is either $ag^2E_5^{{\rm magn}}$
when used in the Hamiltonian simulation or $g^2\hbar (S_4/\hbar)$ when
used in quantum Monte Carlo algorithms), the conjecture
\be
 E_5^{{\rm magn}}/T = S_4/\hbar
\label{MAIN-POINT}
\ee
is literally equivalent to $\: \hbar = aT$ (eq.(1)), as we have argued
earlier on the basis of plasma physics considerations.

\begin{figure}[h]
\begin{center}
\includegraphics[width=0.4\textwidth,angle=-90]{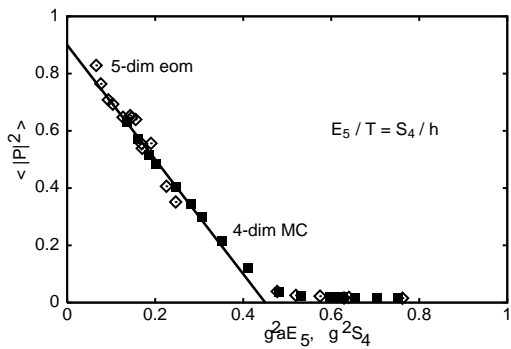}
\end{center}
\caption{ The order parameter, the absolute value square of
the Polyakov line averaged over the lattice and over many
configurations is plotted against the 
4 dimensional plaquette sum in the classical Hamiltonian
(open diamonds) and in the quantum Monte Carlo (full squares)
simulations, respectively.
The scaling of these results coincides if $E_5 / T = S_4/\hbar$.
}
\label{Fig1}
\end{figure}

\va
Now we demonstrate the validity of the relation (\ref{MAIN-POINT})
by numerical computation.
Figure (\ref{Fig1}) shows the absolute value square of the
lattice-averaged Polyakov line (the standard order parameter of
lattice gauge theory) values -- averaged over many quantum Monte Carlo
configurations (full squares) as a function of the 4-dimensional
lattice plaquette sum per plaquette $g^2S_4$. 
On the same plot the same order parameter
is shown as a function of the partial plaquette sum corresponding to
the magnetic energy $ag^2E_5^{{\rm magn}}$
after Hamiltonian
equilibration on the classical 4-dimensional lattice, averaged over
many points alongside a single evolution trajectory at consecutive 
5-th coordinate times (open diamonds). That these two sets of points
belong to the same (in the Coulomb plasma phase linear)
scaling law, proves our main conjecture.

\va

\va
The reason that we do not plot the Polyakov line as usual, as
a function of inverse coupling, is that different couplings
belong to the 4 and to the 5-dimensional simulation once
$\: \hbar = aT$ is valid. In fact the lines fitted to 
our simulation points do not coincide
{\em unless} we assume (\ref{MAIN-POINT}).

\begin{figure}[h]
\begin{center}
\includegraphics[width=0.25\textwidth,angle=-90]{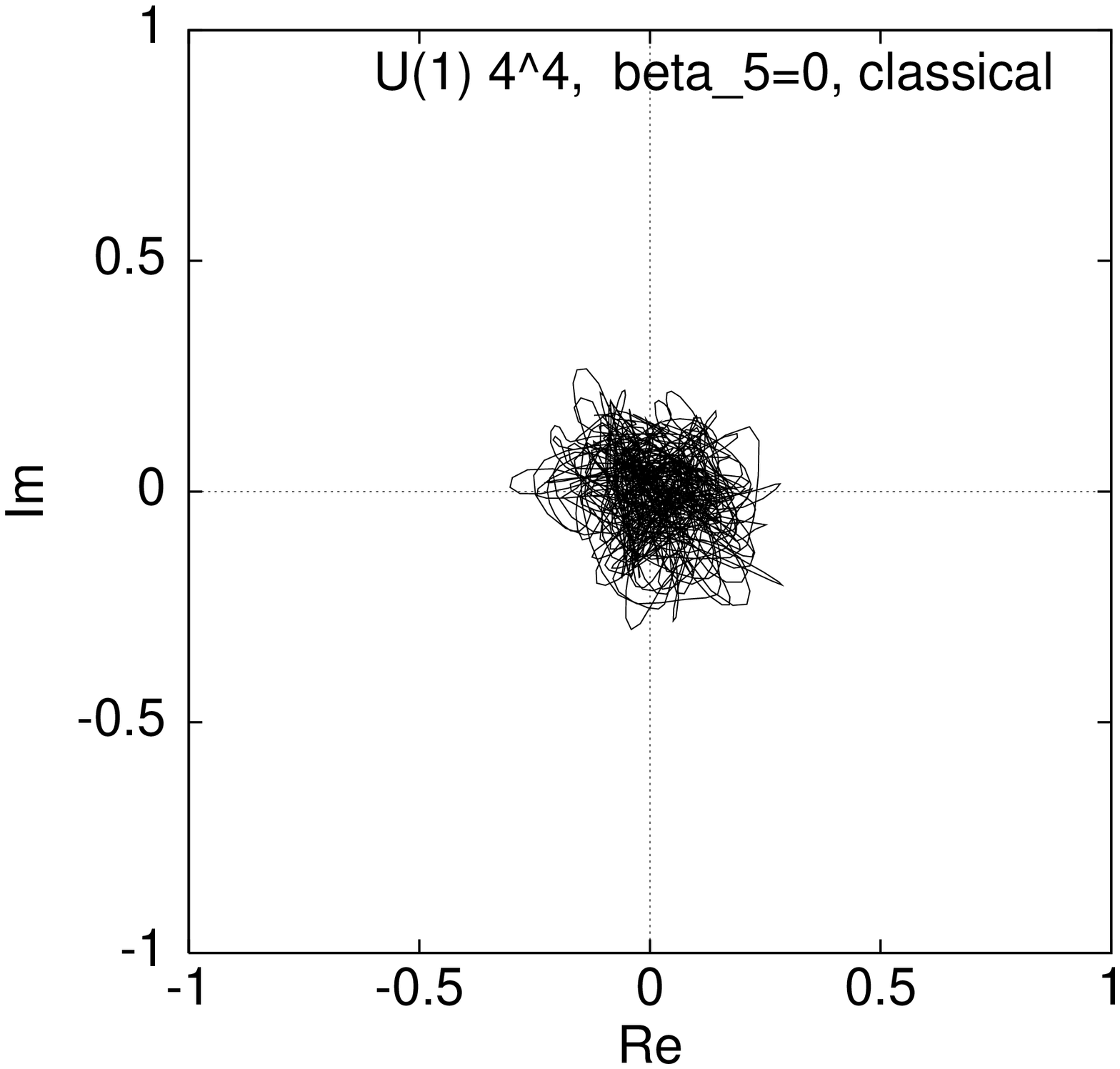}\includegraphics[width=0.25\textwidth,angle=-90]{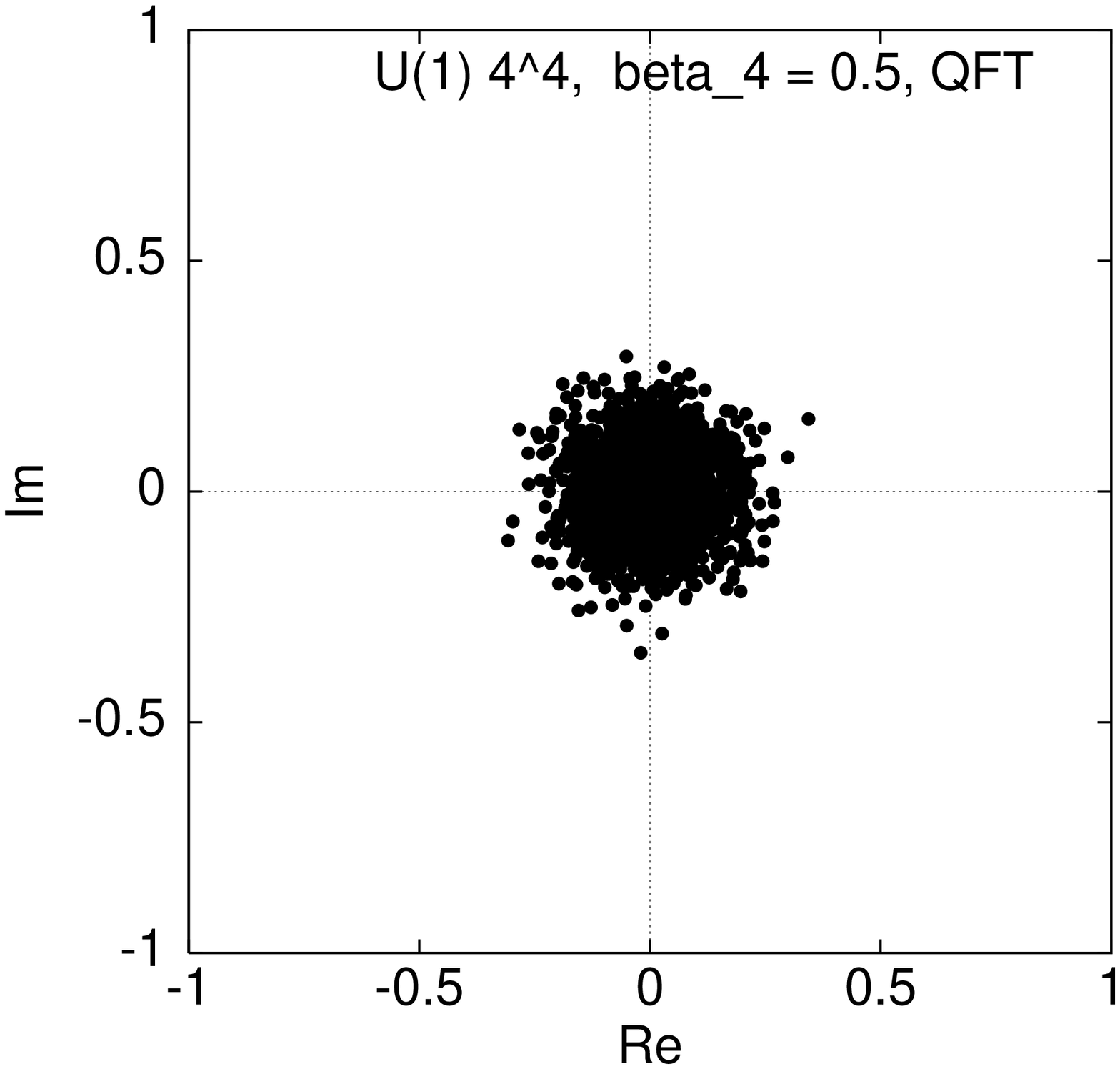}

\includegraphics[width=0.25\textwidth,angle=-90]{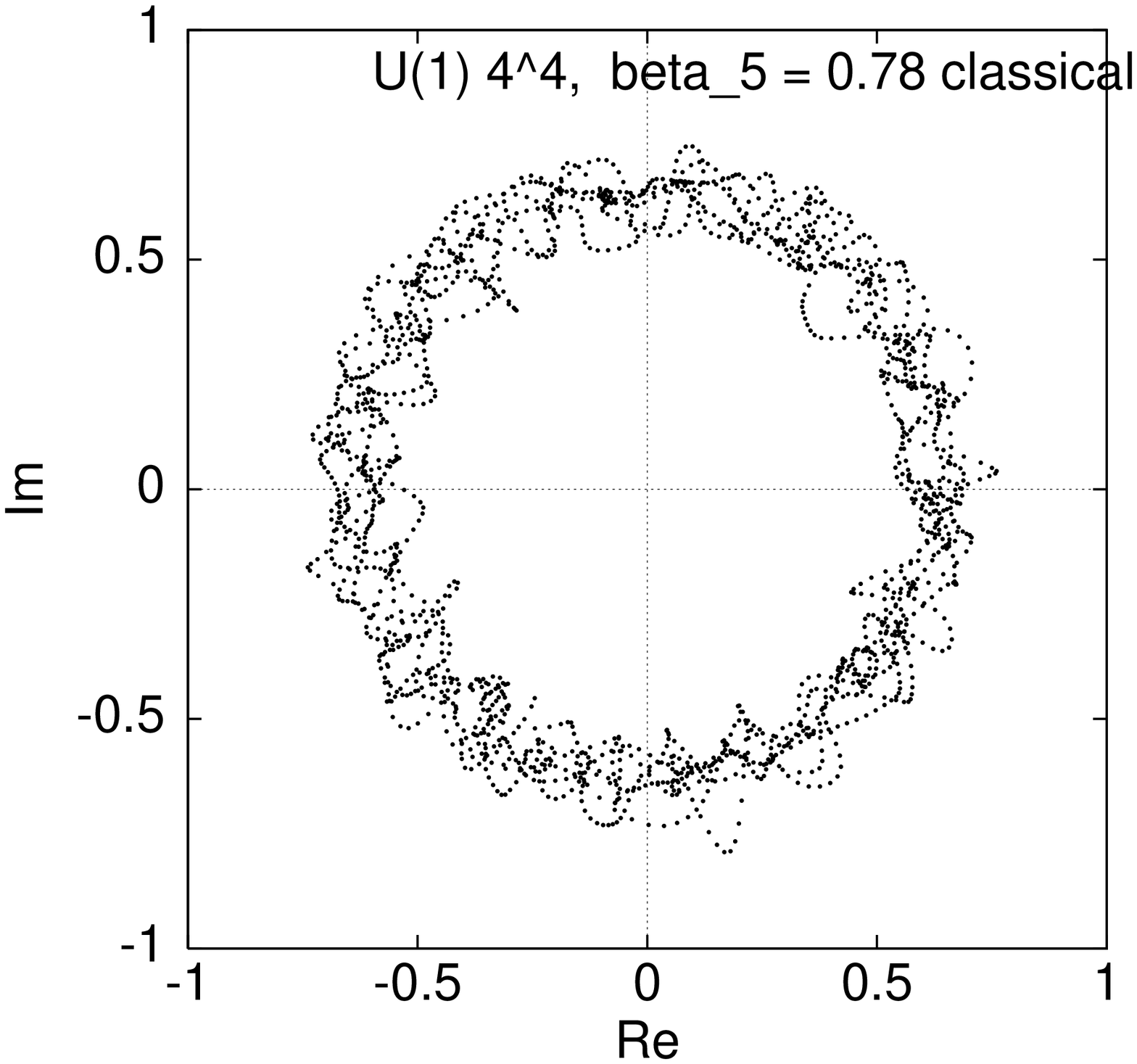}\includegraphics[width=0.25\textwidth,angle=-90]{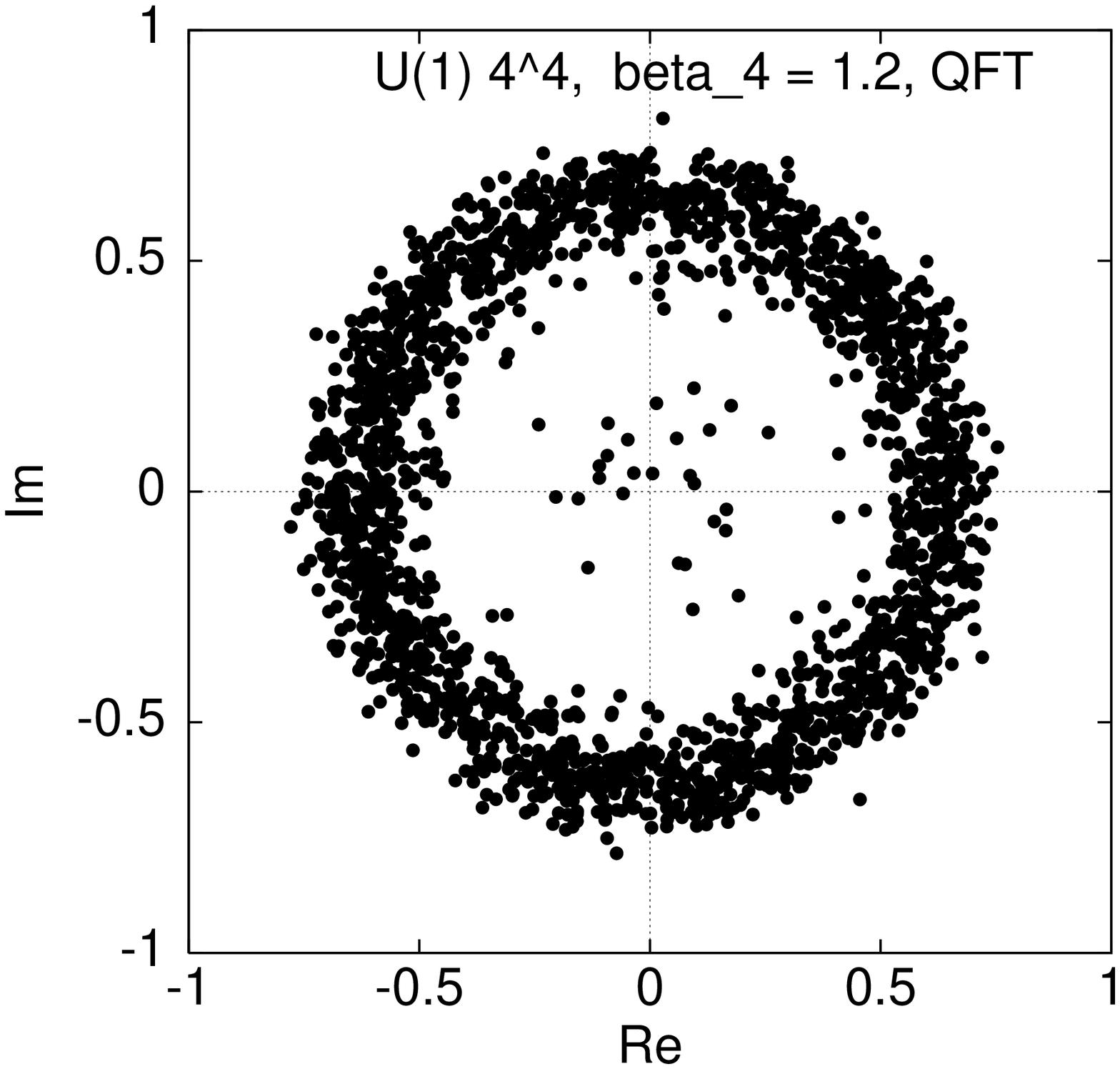}

\end{center}
\caption{
Complex Polyakov line values from 4-dimensional quantum
Monte Carlo simulation (right column) and from 5-dimensional classical
Hamiltonian equation of motion (left column) at $aE_5^{mag} = S_4$.
}
\label{Fig2}
\end{figure}

\va
In order to offer a possibly more direct insight into the
relation of 4 dimensional quantum and 5 dimensional classical
lattice U(1) theories we plot several points on the
complex Polyakov-line plane, both from 4 di\-men\-sion\-al quantum
Monte Carlo (right column) and from 5 dimensional classical
Hamiltonian evolution (left column, actually one single
trajectory is plotted). The left and right parts 
of Fig.(\ref{Fig2}) belong to
different inverse couplings for the 4 and 5-dimensional
cases, the correspondence is made by selecting out pairs of
simulations satisfying (\ref{MAIN-POINT}). 
The 5-dimensional coupling actually does not play any important
role; only the energy content of the configuration is related to it.
Once it is given the equipartition happens due to the very same  
Hamiltonian evolution initially (not shown in the Figure), which
governs the chaotic trajectory covering the same region of configurations
which is generated by quantum Monte Carlo codes.

\va
Now the correspondence between classical and quantum configurations
is excellent, both for the magnitude and for the phase of
Polyakov lines. Some initial points in the middle of the
rings for overcritical couplings $\beta_4 = 1/g_4^2 \ge 1$
are irrelevant; they stem from an initial MC heating phase. The classical
Hamiltonian evolution also had an initial phase equilibrating
electric and magnetic field energy (in 5 dimensions their ratio is
however not 1:1 but 2:3).

\section{Conclusion}\label{concl}

\va
Summarizing we have demonstrated that the mechanism of chaotic 
quantization - conjectured earlier on the basis of non-abelian plasma
physics - works in the practice for lattice gauge theory in
5-dimensional classical form. The correspondence to the
traditional 4-dimensional quantum Monte Carlo simulations
is given by the general formula $\: \hbar = a T$, a formula
encoding physical properties of the higher dimensional lattice 
and field configurations into
the Planck constant. This fact underlines our hope for a
unified classical field theory of gravity and standard particle physics,
in particular for an explanation of standard model parameters
via the mechanism of chaotic quantization, as well as for getting
closer to an insight on the origin of Planck's constant.

\va
Finally we would like to address the question whether factorizing
the Planck constant would not mean to construct a hidden parameter
theory. It is not necessarily the case, since none of the laws
of experimental quantum physics seem to be violated by our results:
the higher dimensional classical dynamics acts as Euclidean 
quantum field theory in 4 dimensions in all respects. 
On the other hand the impossibility of a hidden parameter
is proven for local actions in a strict manner, while the existence
of a higher dimension allows for subtle non-local effects in the
4 dimensions of the physical experience.

\va
Of course, as anything referring to the Planck scale, the theory
of chaotic quantization seems to be speculative at the first glance.
What is -- at least in principle -- better than in the case
of fundamental string theory, that the autocorrelation time-scale,
$\tau \sim \sigma/k^2$ may be effective at scales other, than the
Planck length. From the known experimental fact of the relative
weakness of gravity coupling compared to standard quantum field 
theories (QFT) $g^2_{{\rm grav}} \ll g^2_{{\rm QFT}}$, we conclude
that the time scales beyond which a phenomenon occurs to be quantum,
and for shorter time observations not, separates gravity from the
rest of standard model:
$$ \tau_{{\rm grav}} \ll \tau_{{\rm observ}} \ll \tau_{{\rm QFT}}. $$
As a consequence gravity behaves classically while the other three
known interactions according to quantum field theory.

 
\section*{Acknowledgment}

This work has been supported by the Hungarian National Research
Fund under the contract T-034269.
 


\vfill\eject

\end{document}